\documentstyle[12pt]{article}
\newcommand{\EQ}{\begin{equation}}
\newcommand{\EN}{\end{equation}}
\newcommand{\A}{\begin{array}}
\newcommand{\E}{\end{array}}
\newcommand{\EA}{\begin{eqnarray}}
\newcommand{\EE}{\end{eqnarray}}

\newcommand{\goto}{\rightarrow}
\newcommand{\hs}{\hspace{1mm}}

\newcommand{\D}{{\rm d}}

\newcommand{\th}{\theta}
\newcommand{\siml}{\raisebox{-.6ex}{$\stackrel{<}{\displaystyle{\sim}}$}}
\newcommand{\simg}{\raisebox{-.6ex}{$\stackrel{>}{\displaystyle{\sim}}$}}

\newcommand{\NP}{Nucl. Phys. }

\begin{document}
\setcounter{page}{0}
\topmargin 0pt
\oddsidemargin 5mm
\renewcommand{\thefootnote}{\fnsymbol{footnote}}
\newpage
\setcounter{page}{0}
\begin{titlepage}
\begin{flushright}
January 1992
\end{flushright}
\vspace{1.2cm}
\begin{center}
{\large {\bf EXACT CRITICAL EXPONENTS OF THE STAIRCASE MODEL}}

\vspace{1.9cm}
{\large
Michael L\"{a}ssig} \footnote{
Electronic mail: iff299@DJUKFA11, lassig@iff011.dnet.kfa-juelich.de
}

\vspace{1cm}
{\em Institut f\"ur Festk\"orperforschung \\
     Forschungszentum J\"ulich \\
     5170 J\"ulich, Germany } \\
\end{center}
\vspace{1.1cm}

\setcounter{footnote}{0}

\begin{abstract}
The staircase model is a recently discovered \cite{AlZam.staircase}
one-parameter family of integrable two-dimensional continuum field theories.
We analyze the novel critical behavior of this model, seen as a perturbation of
a minimal conformal theory $M_p$: the leading thermodynamic singularities are
simultaneously governed by all fixed points $M_p, M_{p-1}, \dots, M_3$. The
exponents of the magnetic susceptibility and the specific heat are obtained
exactly. Various corrections to scaling are discussed, among them a new type
specific to crossover phenomena between critical fixed points.
\end{abstract}

\end{titlepage}

\newpage
\renewcommand{\thefootnote}{\arabic{footnote}}

In the past few years, there has been renewed interest in exploring the
consequences  of integrability in two-dimensional statistical systems. At a
critical point, an infinite number of integrals of motion appears if the theory
is conformally invariant; one has even a partial classification of such
universality classes \cite{BPZ}. For some perturbations away from criticality,
a subset of these integrals survives and makes the theory solvable even at
finite correlation length $\xi$. Two types of such systems with a generic
$(p-1)$critical point ($p=3,4,\dots $) described  by the minimal conformal
theory $M_p$  have been known:   \newline
(a) lattice models \cite{ABF}, whose manifold of integrability is parametrized
about the fixed point $M_p$ by a relevant temperature-like  thermodynamic
parameter as well as marginal and irrelevant parameters governing the lattice
effects, and  \newline
(b) exact factorizable scattering matrices for certain relevant perturbations
of the conformal fixed points $M_p$ \cite{Zam.int}. These define massive
continuum  field theories that describe universal scaling behavior.
The thermodynamic Bethe ansatz \cite{AlZam.Potts} is a way to calculate  the
universal ground state energy of the associated Hamiltonian on a circle of
circumference $R$,
\EQ
E_0 (R,\xi) =  \frac{2 \pi}{R} f(R/\xi) \hs.
\EN
The scaling function $f(\rho)$ shows a simple crossover from the thermodynamic
regime $R \gg \xi$ to the conformal regime $R \ll \xi$; its ultraviolet limit
is determined by the central charge $c$ of the asymptotic conformal theory:
$\lim_{\rho \goto 0} f(\rho) = -c/12$.

The ``staircase model" is  a one-parameter family $M(\th_0)$  $(\th_0 > 0)$ of
factorizable scattering theories with a new and more intricate scaling
behavior,
discovered very recently by Al.B. Zamolodchikov \cite{AlZam.staircase}. These
theories contain a single type of massive particles that are characterized by
the purely elastic $S$-matrix \linebreak
$ S(\theta,\th_0) =
(\sinh \theta - i \cosh 2 \th_0)/(\sinh \theta + i \cosh 2 \th_0) $,
written in terms of the Lorentz-invariant rapidity difference $\theta$. The
scaling function $f(\rho, \th_0)$ shows a staircase pattern that interpolates
between all central charges $c_p$ (see fig. 1).  Hence the theory $M(\th_0)$ is
described by a renormalization group trajectory that comes close to each
fixed point $M_p$ for a RG ``time'' interval $\th_0$, whereafter it crosses
over to the next lower fixed point $M_{p-1}$ \cite{AlZam.staircase}.

In another paper \cite{STAIRCASE1}, I argued that from a Lagrangian point of
view, this model can be understood as any fixed point theory $M_p$, perturbed
by a linear combination of its weakest relevant scaling field $\phi_{(1,3)}$
and
its leading irrelevant scaling field  $\phi_{(3,1)}$,
\EQ
{\cal L} = {\cal L}^{\star}_p + t_p \phi_{(1,3)} - \bar t_p \phi_{(3,1)} \hs,
\label{L}
\EN
$t_p > 0$ and $\bar t_p > 0$ are dimensionful coupling constants with
renormalization group eigenvalues $y_p = 4/(p+1)$ and $\bar y_p = - 4/p$,
respectively. The purpose of this Letter is to show in detail that this leads
to a unique critical behavior as a function of the relevant ``temperature"
$t_p$: for $t_p =0 $, the theory is in the universality class of $M_p$, but the
thermodynamic singularities as $t_p \! \searrow \! 0$ are determined not by
$M_p$ alone, but simultaneously by all fixed points $M_p, M_{p-1}, \dots, M_3$.
Specifically, I obtain the exact exponents of the magnetic susceptibility
\EQ
\chi (t_p) \sim t_p ^{-\gamma_p}
\EN
and the specific heat
\EQ
{\cal C} (t_p) \sim t_p^{-\alpha_p} \log t_p
\label{C}
\EN
and discuss various corrections to scaling.

Consider first a crossover between two critical fixed points which is
characterized by a renormalization group trajectory $u(\tau)$. The two-point
correlation functions
$G_{ij} (r,u) \equiv \langle \phi_i (0,u) \phi_j (r,u) \rangle$
of local fields $\phi_i $ are path-ordered integrals
\EA
\lefteqn{G_{ij}({\rm e}^{\tau_2 - \tau_1} r, u(\tau_1)) = } \nonumber
\\ & &
G_{kl} (r, u(\tau_2))
\left ( P {\rm exp}(- \int_{\tau_1}^{\tau_2} \!\D \tau {\bf x} (u(\tau)))
\right )^{\! k}_{\! i}
\left ( P {\rm exp}(- \int_{\tau_1}^{\tau_2} \!\D \tau {\bf x} (u(\tau)))
\right )^{\! l}_{\! j}
\label{CS}
\EE
over the matrix of anomalous dimensions ${\bf x}(u)$, which is given in
perturbation theory about the ultraviolet fixed point by
\EQ
x_i^{\hs j} (u) = x_i^{\hs j} (0) + C_{i\hs k}^{\hs j} u^k + O(u^2)
\EN
in terms of the asymptotic anomalous dimensions and operator product
coefficients $C_{i\hs k}^{\hs j}$. These equations imply the {\em mixing of
fields} under the renormalization group: the infrared asymptotic scaling
fields, i.e. the eigenvectors of the matrix ${\bf x}(u^{\star})$, are linear
combinations of the ultraviolet scaling fields, i.e. the eigenvectors of ${\bf
x}(0)$. If e.g. the  local magnetization $\sigma(r)$ is expanded in the
infrared scaling basis, the lowest appearing eigenvalue $x$ gives its scaling
dimension and the other eigenvalues $x',\dots$ contribute corrections to
scaling. Even if $\sigma(r)$ is chosen a pure scaling field in the ultraviolet,
the mixing of fields dictates that the  infrared asymptotics of its two-point
function $G_{\sigma}(r)$ is $G_{\sigma}(r) \sim r^{-2x} (1 + A r^{-2(x'-x)} +
\dots)$. Besides the usual corrections to scaling due to irrelevant variables
and the nonlinearity of scaling variables, this new type of corrections appears
in the crossover scaling functions between two critical points since the matrix
of anomalous dimensions cannot be diagonalized simultaneously in the
ultraviolet and the infrared. The mixing is, of course, restricted by the
symmetries  that are preserved under the crossover, such as spin-reversal
symmetry or self-duality.

A multiple crossover involving the fixed points $M_p, M_{p-1}, \dots, M_3$ is
characterized by the appearance of $p-1$ different length scales  $\xi_{p+1,p},
\xi_{p,p-1}, \dots, \xi_{3,2} \equiv \xi$. At distances  $\xi_{p'+1,p'} \siml r
\siml \xi_{p',p'-1}$ (corresponding to RG times  $\tau_{p'+1,p'} \siml \tau
\siml \tau_{p',p'-1}$), the behavior of correlation functions is governed by
the fixed point $M_{p'}$; for $r \simg \xi$, they decay exponentially. At any
such fixed point,  the the expansion of the local magnetization contains the
most  relevant scaling field $\phi_{(2,2)}$  of dimension $x_{p'} = 3/(2 p'
(p'+1))$ (the flow of subleading operators is more  subtle due to operator
mixing). Hence, in order to extract the leading thermodynamic  singularity, one
may integrate in Eq. (\ref{CS}) over the piecewise constant  function $x(\tau)=
x_{p'}$ for $\tau_{p'+1,p'} < \tau < \tau_{p',p'-1}$; this suppresses in
addition the familiar corrections to scaling of the form  $(1 + B r^{y_{\rm
i}})$ due to an irrelevant coupling of dimension $y_i$. The singular part of
the  susceptibility  then consists of a sum of terms
\EA
\lefteqn{\int_{\xi_{p'+1,p'}}^{\xi_{p',p'-1}} 2 \pi r \D r G_{\sigma}(r) =
G_{\sigma} (\xi_{p'+1,p'})
\left ( \frac{\xi_{p',p'-1}}{\xi_{p'+1,p'}} \right )^{-2 x_{p'}}
\xi_{p',p'-1}^2 =
\nonumber } \\ & &
\xi_{p,p-1}^{2 - 2 x_p}
\left ( \frac{\xi_{p-1,p-2}}{\xi_{p, p-1}}  \right )^{2 - 2 x_{p-1}} \dots
\left ( \frac{\xi_{p',p'-1}}{\xi_{p'+1,p'}} \right )^{2 - 2 x_{p'}} \hs.
\label{int}
\EE
Since $x_{p'} \leq 1$ for all $p'$, the leading singularity comes from the
integration region $\xi_{4,3} < r < \xi$; the other terms are a third source of
corrections to scaling.

What makes this equation useful is the fact that any two subsequent crossover
length scales have the same ratio
\EQ
\xi_{p',p'-1} / \xi_{p'+1,p'} = {\rm e}^{\th_0} \hs.
\label{ratio}
\EN

This feature of the exact solution is consistent with the Lagrangian
formulation (\ref{L}), as can be inferred from renormalization group and
scaling
arguments (for details see  \cite{STAIRCASE1}). Under the flow between $M_p$
and $M_{p-1}$, the irrelevant  running coupling $\bar u_p$ and the relevant
coupling $u_{p-1}$ mix, which can be expressed as the dimensionless equation
\EQ
t_{p-1} \bar t_{p-1}^{- y_{p-1} / \bar y_{p-1} } =
\bar t_p         t_p^{- \bar y_p / y_p } \hs.
\EN
Expressing these dimensionful couplings  in terms of the crossover length
scales
\EQ
t_{p'}^{-1/y_{p'}} \sim \xi_{p',p'-1} \sim \bar t_{p'-1}^{-1/\bar y_{p'-1}}
\label{xi}
\EN
and using the exponent relation $ y_{p-1} = 4/p = - \bar y_p $ then shows that
the ratio  \linebreak
$(\xi_{p+1,p} / \xi_{p,p-1}) / (\xi_{p,p-1} / \xi_{p-1,p-2})$
is asymptotically constant.

{}From (\ref{int}), (\ref{ratio}) and (\ref{xi}),  one obtains
\EQ
\gamma_p =
\frac{1}{y_p} (2 - 2 x_p + 2 - 2 x_{p-1} + \dots + 2 - 2 x_3 ) =
\frac{(p-2)(2p+1)}{4} \hs.
\label{gamma}
\EN
The local energy density contains at any fixed point the most relevant field
that is even under spin reversal, i.e. $\phi_{(3,3)}$ of  dimension  $\tilde
x_{p'} = 4/(p'(p'+1))$ for $p'>3$ and $\phi_{(1,3)}$ of dimension $\tilde x_3 =
1$ for $p=3$. Hence for the specific heat, one obtains in a similar  way
\EQ
\alpha_p = \frac{1}{y_p}
(2 - 2 \tilde x_p + 2 - 2 \tilde x_{p-1} + \dots + 2 - 2 \tilde x_3 ) =
\frac{p(p-3)}{2} \hs;
\label{alpha}
\EN
the additional logarithmic singularity in Eq. (\ref{C}) comes from the
integration region  $\xi_{4,3} < r < \xi$, which is governed by the Ising fixed
point $M_3$.  Zamolodchikov's $S$-matrix has also been related to multiple
crossovers in the  $D$-series of minimal models \cite{KlassenMelzer.Dflow},
based on a  thermodynamic Bethe ansatz analysis for antiperiodic boundary
conditions  \cite{Fendley.exc}. A $D$-series staircase model has the same
Lagrangian description (\ref{L}) and the same flow of scaling fields that are
even under spin reversal, but Eq. (\ref{gamma}) is replaced by a similar sum
over scaling dimensions of odd $D$-series fields.

{}From a renormalization group point of view, two (related) aspects of this
system are remarkable. \newline
(a) The {\em irrelevant} parameter $\bar t_p$ in (\ref{L}) affects the
{\em leading} exponents associated to the relevant perturbation $t_p$. This
effect, encountered also \cite{Huse} in regime IV of the integrable lattice
models of Andrews, Baxter and Forrester \cite{ABF}, is due to mixing under the
renormalization group \cite{STAIRCASE1}. \newline
(b) Several fixed points democratically share the responsibility for the
exponents (\ref{gamma}) and (\ref{alpha}). Generically, the ratio of different
crossover ``times'' $\tau_{p',p'-1} - \tau_{p'+1,p'}$ would be singular as
$t_p \goto 0$. That this is not the case in the staircase model is yet another
of the amazing fine-tunings occurring in integrable systems.


\newpage

\vspace*{0.7\textheight}
\noindent Fig. 1. The ground state scaling function $f(\rho,\th_0)$
\cite{AlZam.staircase} for the staircase model $M(\th_0)$. All steps have the
same logarithmic width $\th_0$.

\end{document}